\documentclass[amsmath,amssymb,aps,pra,reprint,groupedaddress,showpacs]{revtex4-1}

\usepackage{multirow}
\usepackage{verbatim}
\usepackage{graphicx}
\usepackage{dcolumn}
\usepackage{bm}
\usepackage{longtable}

\newcommand{\ket}[1]{\left\vert #1 \right\rangle}

\newtheorem{theorem*}{Theorem}
\newtheorem{theorem}{Theorem}

\newtheorem{proposition}[theorem]{Proposition}

\begin{document}

\title{Instantaneous Quantum Channel Estimation during Quantum Information Processing}

\author{Yuichiro Fujiwara}
\email[]{yuichiro.fujiwara@caltech.edu}
\affiliation{Division of Physics, Mathematics and Astronomy, California Institute of Technology, MC 253-37, Pasadena, California 91125, USA}

\date{\today}

\begin{abstract}
We present a nonintrusive method for reliably estimating the noise level
during quantum computation and quantum communication protected by quantum error-correcting codes.
As preprocessing of quantum error correction, our scheme estimates the current noise level
through a negligible amount of classical computation using error syndromes
and updates the decoder's knowledge on the spot before inferring the locations of errors.
This preprocessing requires no additional quantum interaction or modification in the system.
The estimate can be of higher quality than the maximum likelihood estimate based on perfect knowledge of channel parameters,
thereby eliminating the need of the unrealistic assumption that the decoder accurately knows channel parameters a priori.
Simulations demonstrate that not only can the decoder pick up on a change of channel parameters,
but even if the channel stays the same, a quantum low-density parity-check code can perform better
when the decoder exploits the on-the-spot estimates instead of the true error probabilities of the quantum channel.
\end{abstract}


\maketitle

\section{Introduction}
Protecting quantum information from noise is of paramount importance to quantum information processing
because \textit{qubits}, the information carriers, are fragile.
\textit{Quantum error-correcting codes} are schemes that encode quantum information into physical qubits
in such a way that errors can be corrected \cite{Lidar:2013}.

Before implementing and performing quantum error correction,
ideally we would like a very accurate channel model, that is, prior knowledge about how errors would manifest on qubits.
The procedure of identifying the behavior of a noisy channel is called \textit{channel estimation} or \textit{parameter estimation} \cite{Fujiwara:2001}.
The known methods employ the technique called \textit{quantum process tomography} \cite{Mohseni:2008},
where probe qubits are fed to the channel and then the behavior of the channel is estimated from the outcome.
In the context of fighting against noise, this standard approach allows for identifying what kind of error would occur and how frequently.

To correct any kind of error in a typical general channel model,
one should be able to correct two types of errors,
namely \textit{bit flips} caused by Pauli operator $X$ and \textit{phase flips} caused by another kind of Pauli operator $Z$ \cite{Nielsen:2000}.
In a situation where the channel introduces a bit flip and/or phase flip on each qubit independently with certain probabilities,
which we assume in most of this paper,
properly implemented parameter estimation may quite accurately reveal the probability $p_X$ that the channel introduces the $X$ error on each qubit
and the same statistical information $p_Z$ about the $Z$ error.
In an ideal situation, one would gain the true values of $p_X$ and $p_Z$, or \textit{perfect knowledge} of the channel parameters.

When constructing a quantum information processing apparatus,
we need to accurately estimate the channel parameters for each component so that
we can install an appropriate quantum error-correcting code tailored to the identified channel behavior.
However, even if we assume that channel parameters can be estimated with no error at the time of constructing an apparatus,
there remain important problems in channel estimation for quantum error correction.

A trivial issue is that the known estimation methods can not be completed instantly on an apparatus that is currently operating.
For instance, this means that any behavioral change in error pattern or frequency during quantum information processing
can not be detected on the spot before attempting quantum error correction.

A sutler but equally critical problem is that perfect knowledge of channel parameters such as the true value of $p_X$ is
not the most helpful side information for quantum error correction once it goes into operation.
In other words, while accurate knowledge of parameters is necessary to choose the right quantum error-correcting code,
it is not the right information for realizing the full potential of the chosen quantum error-correcting code.

To make the latter point clearer, consider the following simple error correction for classical data transmission.
The sender transmits binary information represented by $0$'s and $1$'s.
Assume that the channel flips the symbol of each bit with probability, say, $p=\frac{1}{4}$.
The simplest error correction scheme is to send the same symbol multiple times.
Assume that the sender transmits $5$ copies of each bit in a row so that $0$ is sent as $00000$ and $1$ is encoded as $11111$.
If the receiver knows the error probability $p=\frac{1}{4}<\frac{1}{2}$, the most logical way to infer the correct symbol is by majority vote.
For instance, if the received message is $01000$, the correct message is most likely $0$.

Is the true value of the error probability $p$ the most useful information to this simple error-correcting code?
The answer is no. Excluding the actual error locations which are assumed to be unknown,
it is the actual number of errors that is most useful.
If the number of errors is less than or equal to $2$, both knowing $p$ and knowing the number of errors will lead to the correct guess.
If there are more errors, however, the most logical inference based on perfect knowledge of $p$ fails to reach the correct message.
Yet, knowing the actual number of errors always correctly reveals the original message.
For instance, if the receiver is told that $3$ bits are flipped when $11010$ is received, the most logical inference is $00000$, which is trivially correct.

In general, from the viewpoint of the receiver, the actual number of erroneous bits, or equivalently the \textit{current noise level},
is more useful side information than perfect knowledge of error probability $p$
because the channel parameter $p$ only tells what the noise level is on average.
Therefore, it is natural to ask whether quantum error correction can also exploit knowledge of the current noise level
and, if positive, whether it is possible to estimate it on the spot before inferring errors during quantum information processing.
This paper answers both questions in the affirmative.

We show that it is possible to reliably estimate the number of errors on encoded qubits
without disturbing the quantum state in such a way that no additional quantum circuit or quantum interaction is required.
The estimate is obtained instantaneously through a negligible amount of classical computation
before the decoder of the quantum error-correcting code starts to infer the types and locations of errors.
In other words, the current noise level can be estimated as preprocessing of quantum error correction at virtually no cost.

The estimate can be fed into the decoder each time to make quantum error correction more reliable
by letting it adaptively respond to the current noise level.
This means that not only can the decoder pick up on a change of a channel parameter,
but even if $p_X$ and $p_Z$ stay exactly the same,
it can also follow the natural deviations from the expected number of errors to the extent the accuracy and precision of estimation allows.

It is shown that our instantaneous quantum channel estimation can be implemented with a quantum error-correcting code
that can take advantage of the \textit{sum-product algorithm} \cite{MacKay:2003,MacKay:2004},
which is among the most sophisticated and popular decoding methods available in coding theory.
Simulations demonstrate that the on-the-spot estimate can be of very high quality to the extent that
the decoder no longer needs perfect knowledge of channel parameters during quantum error correction.

\section{Preliminaries}
Our estimation scheme is integrated with quantum error correction that takes advantage of classical error correction.
For the basic notions and facts in classical coding theory and quantum error correction,
we refer the reader to standard textbooks such as \cite{MacKay:2003,Lidar:2013,Nielsen:2000}.

A \textit{binary linear} $[n,k]$ \textit{code} is a $k$-dimensional subspace $\mathcal{C}$ of the $n$-dimensional vector space $\mathbb{F}_2^n$
over the finite field $\mathbb{F}_2$ with exactly two elements $\{0, 1\}$,
so that $\mathcal{C}$ encodes $k$ bits of information into $n$ physical bits as a classical error-correcting code.
The \textit{dual code} $\mathcal{C}^\perp$ of $\mathcal{C}$ is defined as
$\mathcal{C}^\perp=\{\boldsymbol{d}\in\mathbb{F}_2^n\mid\boldsymbol{c}\cdot\boldsymbol{d}=\boldsymbol{0}\mbox{\ for any\ }\boldsymbol{c}\in\mathcal{C}\}$.

A binary linear code $\mathcal{C}$ can be seen as the null space $\{\boldsymbol{c}\in\mathbb{F}_2^n\mid H\boldsymbol{c}=\boldsymbol{0}\}$
of some $(n-k)\times n$ matrix $H$ over $\mathbb{F}_2$. The matrix $H$ is called a \textit{parity-check matrix} of $\mathcal{C}$.
Take an $n$-dimensional vector $\boldsymbol{e}=(e_0,\dots,e_{n-1})\in\mathbb{F}_2^n$.
Assume that a message $\boldsymbol{c}=(c_0,\dots,c_{n-1})\in\mathcal{C}$ is sent and that the vector $\boldsymbol{c}+\boldsymbol{e}$ is received,
which means that the bit $c_i$ is flipped by the channel if $e_i = 1$ and it is intact if $e_i = 0$.
The traditional error correction method computes the $k$-dimensional vector $\boldsymbol{s}=H(\boldsymbol{c}+\boldsymbol{e})=H\boldsymbol{e}$,
called the \textit{syndrome}, and then infers $\boldsymbol{e}$
from $\boldsymbol{s}$ with the help of side information such as the error probability $p$.

Similar to the classical case, an $[[n,k]]$ quantum error-correcting code encodes $k$ qubits of quantum information into $n$ physical qubits.
A quantum error-correcting code can be constructed from binary linear codes.
Let $\mathcal{C}_1$, $\mathcal{C}_2$ be a pair of binary linear codes of parameters $[n,k_1]$ and $[n,k_2]$ respectively.
If $\mathcal{C}_1$ contains the dual code $\mathcal{C}_2^\perp$, that is, $\mathcal{C}_2^\perp\subseteq\mathcal{C}_1$,
then an $[[n,k_1+k_2-n]]$ quantum error-correcting code, called
a \textit{Calderbank-Shor-Steane} (CSS) \textit{code} \cite{Calderbank:1996,Steane:1996}, can be constructed.

For a unitary operator $U$ and a binary vector $\boldsymbol{a}=(a_0,\dots,a_{n-1})\in\mathbb{F}_2^n$,
define $U^{\boldsymbol{a}}$ as the $n$-fold tensor product $O_0 \otimes\dots\otimes O_{n-1}$, where
$O_i = U$ if $a_i = 1$ and $O_i$ is the identity operator otherwise.
A CSS code exploits a technique called \textit{discretization} so that
its error detection measurement takes any error operator $E$ introduced by the channel
to a combination of bit flips $X$, phase flips $Z$, and both at the same time.
Let $H_1$ and $H_2$ be parity-check matrices of the binary linear codes $\mathcal{C}_1$ and $\mathcal{C}_2$ such that
$\mathcal{C}_2^\perp\subseteq \mathcal{C}_1$.
Then,  with $2k$ ancilla qubits, appropriate measurement discretizes the error $E$ on an $n$-qubit state $\ket{\psi}$
encoded by a CSS code as follows:
\[E\ket{\psi}\rightarrow\ket{H_1\boldsymbol{e}_X}\ket{H_2\boldsymbol{e}_Z}X^{\boldsymbol{e}_X}Z^{\boldsymbol{e}_Z}\ket{\psi},\]
where $\boldsymbol{e}_X=(e^X_0,\dots,e^X_{n-1})\in\mathbb{F}_2^n$
is the $n$-dimensional vector such that $e^X_i=1$ if a bit flip occurred on the $i$th qubit and $e^X_i=0$ otherwise,
and $\boldsymbol{e}_Z$ is the $n$-dimensional vector representing phase flips the same way.
Measuring ancilla qubits, we obtain the syndrome $H_1\boldsymbol{e}_X$ for bit flips
as a binary $(n-k_1)$-dimensional vector
and the other syndrome $H_2\boldsymbol{e}_Z$ for phase flips as a binary $(n-k_2)$-dimensional vector.
By exploiting the error correction method for binary linear codes,
we may infer $\boldsymbol{e}_X$ and $\boldsymbol{e}_Z$ from the syndromes and side information such as
$p_X$ and $p_Z$ we learned when constructing the apparatus.

What we prove here is that if $H_1$ and $H_2$ are chosen suitably,
a tiny amount of classical computation with the syndromes
can estimate the current noise level, or equivalently the numbers of bit flips and phase flips.
Thus, the next step where the decoder infers $\boldsymbol{e}_X$ and $\boldsymbol{e}_X$
can exploit the estimated current noise level as more useful and updated side information.

\section{Instantaneous noise level estimation}
Now we show how to estimate the current noise level for bit flips from $H_1\boldsymbol{e}_X$
before inferring the error vector $\boldsymbol{e}_X$.
Because the same estimation can be performed for phase flips from $H_1\boldsymbol{e}_Z$,
the rest of this paper focuses on bit flips.
A remark on extensions to other channel models will be given at the end.

We assume that the $X$ operator acts on each qubit independently with probability $p<\frac{1}{2}$.
Hence, in terms of bit flips, the quantum channel is modeled as
a binary symmetric channel with error probability $p$ in the language of coding theory.
The current error probability $p$ is not necessarily equal to the original value $p_X$
because we assume that the frequency of errors may change.
No prior knowledge about the current value is assumed except the assumption that it is less than $\frac{1}{2}$.

Recall that $H_1$ is a parity-check matrix of binary linear $[n,k_1]$ code $\mathcal{C}_1$
with $n-k_1$ linearly independent rows and $n$ columns.
The \textit{weight} $\operatorname{wt}(\boldsymbol{r})$ of a vector $\boldsymbol{r}$ over $\mathbb{F}_2$ is the number of nonzero entries,
that is, the number of $1$'s.
For the sake of simplicity, we assume that every row of $H_1$ is of weight $r$.

Our estimation bases on the following approximation.
\begin{proposition}[\cite{Toto-Zarasoa:2011}]\label{prop1}
Let $\boldsymbol{e}_X$ be a Bernoulli process with $n$ trials and probability $p$,
and $H_1$ an $(n-k_1)\times n$ parity-check matrix of a binary linear $[n,k_1]$ code in which every row is of weight $r$.
The weight of the syndrome $\boldsymbol{s}_1=H_1\boldsymbol{e}_X$ can be approximated by a random variable that follows
the binomial distribution of parameters $n-k_1$ and $q_{r,p}$, where
\begin{align}\label{EQp}
q_{r,p} = \sum_{\substack{1\leq i\leq r\\ i\ \text{odd}}}{{r}\choose{i}}p^i(1-p)^{r-i}.
\end{align}
\end{proposition}

Note that the right-hand side of Equation (\ref{EQp}) can be simplified to $\frac{1-(1-2p)^r}{2}$.
Assuming the approximation given in Proposition \ref{prop1},
the probability $P\left[\operatorname{wt}(\boldsymbol{s}_1)=s\right]$ that the weight of the syndrome is $s$ is
\[P\left[\operatorname{wt}(\boldsymbol{s}_1)=s\right]={{n-k_1}\choose{s}}q_{r,p}^s(1-q_{r,p})^{n-k_1-s}.\]
Hence, given that $\operatorname{wt}(\boldsymbol{s}_1)=s$,
the maximum likelihood estimate $\hat{p}_s$ of $p$ is
\begin{align}\label{MLps}
\hat{p}_s &=\arg\max_x\left\{q_{r,x}^s(1-q_{r,x})^{n-k_1-s}\right\} \notag\\
&=
\begin{cases}
\frac{1}{2}-\frac{1}{2}\left(1-\frac{2s}{n-k_1}\right)^{\frac{1}{r}} & \text{if} \ \frac{s}{n-k_1} \leq \frac{1}{2},\\
\frac{1}{2} & \text{otherwise}.
\end{cases}
\end{align}
The point is that the nearest integer $[n\hat{p}_s]$ is an estimate of the number of bit flips that actually occurred within the encoded $n$-qubit block
because $\hat{p}_s$ is calculated from the Bernoulli process $\boldsymbol{e}_X$ of a single whole set of $n$ trials.
In other words, $\hat{p}_s$ is more strongly correlated with what just happened on qubits at hand
than how the channel behavior would average out over the course of time.

Another important fact is that our estimation scheme does not impose any overhead except the negligibly small calculation
by the closed form given in Equation (\ref{MLps}).
This is because the only necessary information, which is the syndromes, is required regardless by the inference of the types and locations of errors.

The mean $\mu$ of $\hat{p}_{\operatorname{wt}(\boldsymbol{s}_1)}$ is
\[\mu=\frac{1}{2}-\frac{1}{2}\sum_{i = 0}^{\lfloor\frac{m}{2}\rfloor}{{m}\choose{i}}q_{r,x}^i(1-q_{r,x})^{m-i}\left(1-\frac{2i}{m}\right)^{\frac{1}{r}},\]
where $m=n-k_1$.
When seen as an estimator of $p$, the mean squared error $\mathrm{MSE}(\hat{p}_{\operatorname{wt}(\boldsymbol{s}_1)})$ is
\begin{align*}
\mathrm{MSE}(\hat{p}_{\operatorname{wt}(\boldsymbol{s}_1)})&=\mathrm{E}\left[\left(\hat{p}_{\operatorname{wt}(\boldsymbol{s}_1)}-p\right)^2\right]\\
&= p^2-2p\mu+\frac{1}{4}\\
&\quad+\frac{1}{4}\sum_{i = 0}^{\lfloor\frac{m}{2}\rfloor}{{m}\choose{i}}q_{r,x}^i(1-q_{r,x})^{m-i}\\
&\quad\times\left(\left(1-\frac{2i}{m}\right)^{\frac{2}{r}}-2\left(1-\frac{2i}{m}\right)^{\frac{1}{r}}\right).
\end{align*}
In principle, one may exactly compute the more relevant quantity, namely the expected discrepancy
between a realization of $\frac{\operatorname{wt}(\boldsymbol{e}_X)}{n}$ and our estimate,
although it may be computationally infeasible for a parity-check matrix of practical size.

The key to effectively exploiting our scheme is choosing a suitable parity-check matrix
such that the approximation in Proposition \ref{prop1} is reasonable
and such that the estimation is of high quality.
It should be noted that it is not the same as choosing a suitable binary linear code
because one same code has many different parity-check matrices of different column and row weights.

In general, making column weights smaller decreases correlations between bits in a syndrome
and hence improves the accuracy of the approximation in Proposition \ref{prop1}.
By the same token, the number of overlaps of the positions of $1$'s between any pair of rows should be small.
In addition, the row weight $r$ should be relatively small to keep $q_{r,p}$ sensitive to $p$.
Finally, all else being equal, larger $n$ and smaller $k_1$ are desirable
because estimation becomes more reliable as $m=n-k_1$ becomes larger.

Here we illustrate our estimation method through an example case.
The quantum error-correcting code we use exploits a state-of-the-art decoding technique, the sum-product algorithm, for binary linear codes.
A \textit{low-density parity-check} (LDPC) \textit{code} is a linear code that admits a parity-check matrix with a small number of nonzero entries
such that iterative decoding algorithms perform well \cite{MacKay:2003}.
It is known that well-designed LDPC codes can nearly attain the channel capacity, which is the theoretical limit of error correction \cite{Bonello:2011}.
A CSS code that uses LDPC codes as its ingredients is called a \textit{quantum LDPC code}.
Typically, the column weights of a parity-check matrix of an LDPC code are only a few to several.
Row weights are also quite small.
A well-designed LDPC code tends to have a very small number of overlaps of the positions of $1$'s between a pair of rows
because rows with more than one overlap degrades the performance of its decoding algorithm.
In addition, the sum-product algorithm and most of its variations achieve their characteristic excellent error correction performance
by exploiting information about the noise level
(see \cite{MacKay:2003a,Qi:2006,Saeedi:2007,Hagiwara:2012} for the effect of a mismatch between actual and assumed noise levels).
Therefore, quantum LDPC codes with good error correction performance are naturally suited for our purpose.

Among various known construction techniques for quantum LDPC codes,
Construction B given in \cite{MacKay:2004} is one of the most successful ones.
Following Figure 6 of \cite{MacKay:2004}, we set $n=3786$ and $k_1=2366$.
The row and column weights significantly affect the expected performance of an LDPC code of given parameters $n$ and $k_1$ \cite{Richardson:2008}.
We adjusted our parity-check matrix so that the block error rate (BLER) reaches $5\times10^{-5}$ roughly at $p=0.02$ with the sum-product algorithm.
Every row is of weight $24$. The column weights are nearly uniform with the mean weight being $10$.
This LDPC code contains its dual code,
so that the resulting quantum LDPC code is of parameters $[[3786,946]]$.

We simulated $X$ errors by binary symmetric channels and estimated the current noise level each time by Equation (\ref{MLps}).
Table \ref{MSEsim} shows the mean of the squared errors of estimates obtained through simulations.
Note that the maximum likelihood estimate of the current noise level by perfect knowledge of $p$ is $p$ itself.
Hence, the corresponding quality measure for this perfect knowledge estimator is the variance of the binomial distribution divided by $n^2$.
As shown in Table \ref{MSEsim}, our estimates are of higher quality
than the maximum likelihood estimates based on perfect knowledge in terms of expected discrepancy.
\begin{table}[h!t]\caption{Quality of estimates.\label{MSEsim}}
\begin{ruledtabular}
\begin{tabular}{llll}
$\ \ p$&
$\ \operatorname{MSE}(\hat{p})$&
$\ \operatorname{MSE}(p)$&
$n\hat{p}/\operatorname{wt}(\boldsymbol{e}_X)$\\\hline
0.0175 & $1.0\times10^{-6}$ & $4.5\times10^{-6}$ & 1.007\\
0.02 & $1.4\times10^{-6}$ & $5.1\times10^{-6}$ & 1.008\\
0.0225 & $2.0\times10^{-6}$ & $5.8\times10^{-6}$ & 1.008\\
0.025 & $2.8\times10^{-6}$ & $6.4\times10^{-6}$ & 1.009\\
0.0275 & $3.8\times10^{-6}$ & $7.0\times10^{-6}$ & 1.009\\
0.03 & $5.1\times10^{-6}$ & $7.6\times10^{-6}$ & 1.010\\
0.0325 & $6.9\times10^{-6}$ & $8.3\times10^{-6}$ & 1.011
\end{tabular}
\end{ruledtabular}
\end{table}

\begin{figure}[h!t]
\includegraphics[scale=0.63]{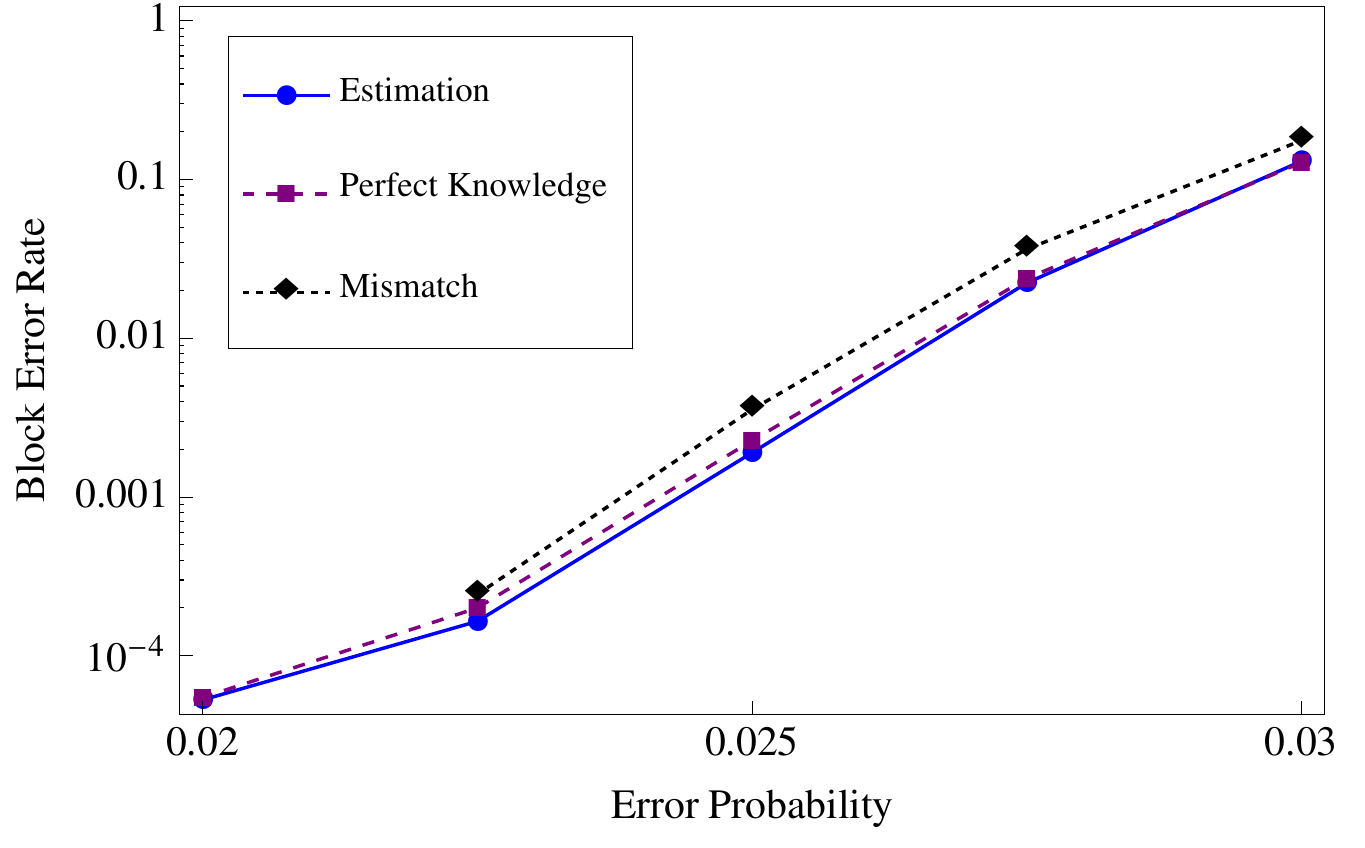}\caption{(Color online) Performance of quantum LDPC code.\label{BLERsim}}
\end{figure}

Figure \ref{BLERsim} plots the BLER $b_p$ of the LDPC code
decoded by the sum-product algorithm over a binary symmetric channel with error probability $p$.
If bit flips and phase flips are treated separately,
the BLER of the corresponding quantum LDPC code achieves $1-(1-b_p)^2 \approx 2b_p$ over a depolarizing channel
with equal error probability $\frac{p}{2}$ for each of the three types of Pauli errors.
We compared three scenarios: the decoder always assumes $p=0.02$ regardless of the actual channel parameter $p$,
which is the most realistic assumption without our method;
the decoder has perfect knowledge of $p$, which is unrealistically optimistic;
and the decoder uses the estimate $\hat{p}_s$ each time.
As illustrated in Figure \ref{BLERsim}, feeding $\hat{p}_s$ to the sum-product decoder completely suppresses the detrimental effect of the mismatch
between actual and assumed noise levels.
In fact, decoding with the estimate $\hat{p}_s$ achieved a better BLER for $p < 0.03$ than with the true value of $p$
because $\hat{p}_s$ is more strongly correlated with the actual number of bit flips than with what is expected on average.
The confidence level of the detection of improvement over perfect knowledge at $p=0.0225$ is $3.9\sigma$.
No advantage over perfect knowledge was observed
when the quantum LDPC code was overwhelmed by too much noise or overkill for too low a noise level.

\section{Concluding remarks}
The instantaneous quantum channel estimation we developed here
makes the unrealistic, ideal assumption of perfect knowledge unnecessary
when appropriate quantum error-correcting codes and their sophisticated decoding algorithms are employed.
In fact, not only does our scheme suppress the negative effect of incorrect knowledge of channel parameters,
but it can also give a better BLER than if perfect knowledge is available
because to the eye of the decoder, the actual number of errors is more relevant than how the channel behaves on average.

It should be noted that the idea presented here is particularly more suited for quantum error correction than classical error correction.
This is because LDPC codes in electrical engineering typically make use of soft information
represented by continuous values instead of binary syndromes.
Since good nonintrusive estimation can be done using soft information in the classical case \cite{Pauluzzi:2000},
it is not as appealing if there is no other reason
to convert soft information into binary data, a process known as a \textit{hard decision} \cite{Lechner:2013}.
Contrary to the classical case, active quantum error correction naturally involves discretization, which is a form of hard decision.
Hence, quantum information processing is exactly the kind of application that benefits from our method.

It is possible to extend our method to different channel models.
For instance, decoherence due to amplitude and phase damping can be approximated by
Pauli operators $X$, $Y$, and $Z$ through Pauli twirling \cite{Silva:2008}.
In this case, we can derive the current noise level for $Y$ from our estimates of $p_X$ and $p_Z$.
Hence, the decoder can also exploit the correlation
between bit flips and phase flips due to the $Y$ operator \cite{MacKay:2004,Wang:2012,Denise:2013}.

An interesting question is how to effectively use knowledge of the current noise level.
While we simply used the estimation for the initialization of the sum-product algorithm
by feeding the estimate as the assumed noise level,
there may be a more sophisticated way to exploit the knowledge.

\end{document}